# MLP Aware Scheduling Techniques in Multithreaded Processors

Murthy Durbhakula

Indian Institute of Technology Hyderabad, India

cs15resch11013@iith.ac.in, murthy.durbhakula@gmail.com

**Abstract:** Major chip manufacturers have all introduced Multithreaded processors. These processors are used for running a variety of workloads. Efficient resource utilization is an important design aspect in such processors. Particularly, it is important to take advantage of available memory-level parallelism(MLP). In this paper I propose a MLP aware operating system (OS) scheduling algorithm for Multithreaded Multi-core processors. By observing the MLP available in each thread and by balancing it with available MLP resources in the system the OS will come up with a new schedule of threads for the next quantum that could potentially improve overall performance. We do a qualitative comparison of our solution with other hardware and software techniques. This work can be extended by doing a quantitative evaluation and by further refining the scheduling optimization.

## 1 Introduction

Major chip manufacturers have all introduced multithreaded multi-core processors. Such chips are used for running various multi-threaded and multi-programmed workloads. Efficient resource utilization is an important design aspect on such processors. It is important to schedule threads in such a way that there is balance of resource utilization. For instance, scheduling multiple memory intensive workloads on the same multi-threaded processor will result in stalls due to unavailability of Miss Status Holding Registers (MSHRs). In this paper I propose Operating System scheduling policies which take memory-level parallelism (MLP) into account while scheduling threads on multithreaded multi-core processors. By observing MLP for each thread in every scheduling quantum the OS will determine where to schedule threads for the next quantum thereby resulting a balanced utilization of resources and an improvement in overall performance. The rest of the paper is organized as follows: Section 2 presents MLP aware OS scheduling optimization. Section 3 briefly describes the qualitative methodology I used in evaluation. Section 4 presents results. Section 5 describes related work and Section 6 presents conclusions.

## 2 OS Scheduling Optimization

The hardware support we need for this OS scheduling optimization are dedicated hardware counters per thread that keep track of average MLP. These metrics are calculated for a time window T which is potentially less than duration of scheduling quantum. This is to capture phase based behavior and not be biased by cummulative/historical behavior over large periods of time. Average MLP for a thread can be calculated as average MSHR occupancy of that thread over a time period T. For the next scheduling quantum the operating system will use these counters and schedule such a way that the cummulative MLP of threads do not exceed the number of MSHRs. In fact if possible it is better to schedule threads such that cummulative MLP is as close to total number of MSHRs as possible. This way we take advantage of maximum MLP the system can support.

Below is the pseudo-code of the optimized scheduling algorithm

Input: Threads T0, T1, ...TN with counters MLP0,MLP1,....MLPN. Current schedule S_current which has mapping of N threads to N slots. For the sake of simplicity we consider a single socket system with K multi-threaded processors that can support L threads per processor. K*L = N.

Output: New schedule S_next with new mapping of threads to nodes

begin

1. Sort all MLP counters in descending order. The complexity of this step is Nlog(N)

2. Walk through sorted list in descending order. Assign each of the top K MLP threads to K multi-threaded processors P0, P1,...PK.  Then take next top K MLP threads and assign them in reverse order to P0, P1, ...PK. This is to balance the MLP allocation to K processors. So highest MLP thread in the second set of K MLP threads will go to PK instead of P0.  Now for the third set of next highest K MLP threads we alternate back to original method assigning highest MLP thread to P0. This way we do alternating assignments until we complete assigning all threads to all K processors. It is possible that in this method we may assign threads to a processor in such a way that we may exceed total MLP supported by a processor. An alternate method in such a scenario would be to assign more than L threads to some processors. However for load balancing reasons we do not do this. We assign L threads to every processor. This can be done in linear time.

end

Overall complexity is O(Nlog(N))

## 3 Methodology

I am using a qualitative methodology to compare the contribution of this paper with other existing approaches. Particularly I compared with hardware solutions and software solutions and the metrics I used are:

i) Need software support: That is, does the approach require changes from software or will it work seamlessly with existing software

ii) Flexibility: That is, can the idea be improved or configured later on. Either software or hardware/software hybrid solutions have this advantage

iii) Verification complexity: Hardware solutions generally have verification complexity. They need to be fully verified before they can ship. Whereas software solutions can be potentially patched.

This work can be extended by doing a quantitative evaluation with various workloads.

## 4 Results

### 4.1 Hardware solution

**Biasing instruction issue to increase MLP**

During instruction issue we can bias the picker to pick load instructions whose addresses are known and can be predicted to be a miss to increase the MLP of the application. For instance if we know that there is a cache miss pending to line A and next cache line A+1 and if we see load requests to line A+2 and line A+3 pending in the load queue then it's better to issue them as soon as possible as there is a likely chance that they too are cache misses.

### 4.2 Software solutions

**Compiler optimizations**

Researchers in the past have proposed compiler optimizations/code transformations [1] to improve the MLP in an application. Our work is orthogonal in the sense that it can also be applied to such compiler optimized workloads. This is a software solution which has an advantage that it is flexible. However in order for existing legacy applications to take advantage of this solution they need to be recompiled, which can be a drawback particularly if source code is not available.

| Solution | Need software changes | Flexibility | Hardware Verification complexity |
|---|---|---|---|
| MLP Aware OS Scheduling | Yes | Yes | No |
| Biasing Instruction Issue | No | No | Yes |
| Compiler Optimizations | Yes | Yes | No |

**Table 1: Comparison of Various Solutions**

## 5 Related work

I have already discussed some related work in the previous section. In this section I am going to discuss some more. Craeynest et al [2] has proposed using run ahead execution to exploit far-distance MLP in multithreaded processors. Whereas in our work we propose using OS scheduling techniques to maximize utilization of available MLP with little hardware support.

Tang et al [3] has proposed compiler techniques to co-optimize both memory-level and cache-level parallelism for last-level-caches. Whereas in our work we propose using run-time information to optimize for MLP.

Mutlu et al [4] proposed optimizing DRAM scheduling techniques to improve intra-thread bank-level parallelism while preserving row buffer locality. This is a pure hardware technique. Whereas in our work we propose using OS scheduling optimization to maximize utilization of available MLP with little hardware support.

## 6 Conclusions

Major chip manufacturers have all introduced multithreaded multi-core processors. Such chips are used for running various multi-threaded and multi-programmed workloads. Efficient resource utilization is an important design aspect on such processors. In this paper I have presented an OS scheduling optimization that keeps track of MLP available in every thread and uses that information to optimally schedule threads for subsequent scheduling quanta for maximizing resource utilization as well as overall performance.

## References

1. Vijay S. Pai and Sarita Adve. "Code Transformations to Improve Memory Parallelism." In Proceedings of the 32nd Annual ACM/IEEE International Symposium on Microarchitecture. 1999

2. Kenzo Van Craeynest, Stijn Eyerman, and Lieven Eeckhout. "MLP-Aware Runahead Threads in a Simultaneous Multithreading Processor." Proceedings of the 4th International Conference on High Performance Embedded Architectures and Compilers. 2008.

3. Xulong Tang, Mahmut Kandemir, Mustafa Karakov, and Meenakshi Arunachalam. "Co-optimizing memory-level parallelism and cache-level parallelism." Proceedings of the 40th ACM SIGPLAN Conference on Programming Language Design and Implementation. 2019.

4. Onur Mutlu and Thomas Moscibroda. "Parallelism-Aware Batch Scheduling: Enhancing Both Performance and Fairness of Shared DRAM Systems." In Proceedings of the 35th Annual International Symposium on Computer Architecture. 2008.